\documentclass[fleqn,twoside]{article}
\usepackage{espcrc2}

\usepackage{graphicx}

\usepackage{amssymb}

\usepackage[figuresright]{rotating}

\newcommand{\AmS}{{\protect\the\textfont2
  A\kern-.1667em\lower.5ex\hbox{M}\kern-.125emS}}

\title{Finite temperature 2-color QCD for real and imaginary 
chemical potential}

\author{Pietro Giudice\address[CS]{Dipartimento di Fisica, 
Universit\`a della Calabria \\
\& Istituto Nazionale di Fisica Nucleare, Gruppo Collegato di Cosenza, Italy}
and Alessandro Papa\addressmark[CS]}
       
\begin{document}

\begin{abstract}
In this work we study the finite temperature SU(2) gauge theory with 
staggered fermions for real and imaginary chemical potential. We test the 
method of analytical continuation of Monte Carlo results obtained for 
imaginary chemical potential, by comparison  with those obtained 
{\em directly} for real chemical potential.
\vspace{1pc}
\end{abstract}

\maketitle

\section{INTRODUCTION}
Understanding the QCD phase diagram in the temperature--chemical potential
$(T,\mu)$ plane is important for the implications in cosmology,  
astrophysics and heavy-ion collisions, where quark--gluon 
plasma could be formed. The main purpose of the most recent studies is to 
localize the critical lines in the phase diagram and to determine the kind of 
transition across them.
The most powerful tool for such investigation is the formulation of the    
theory on a space--time lattice. Unfortunately, at nonzero chemical potential 
the determinant of the fermion matrix becomes complex, thus requesting 
nonstandard approaches to Monte Carlo simulations. 

One way to circumvent this problem is based on the idea to perform numerical 
simulations at {\em imaginary} chemical potential~\cite{AKW99} $\mu=i\mu_I$ 
and to analytically continue~\cite{Lom00,HLP01,DLa,Pd} the results 
to real $\mu$. The advantage is that for imaginary $\mu$ the fermion determinant 
is again real, so simulations are feasible in QCD.  Moreover, this method allows to 
check the results obtained by other approaches.
On the other side the method is applicable as long as one remains inside a 
region of analyticity of the partition function in the $(T,\mu_I)$ 
plane.

\section{PHASE DIAGRAM}
Roberge and Weiss (RW) have shown~\cite{RW86} that the partition function of 
the SU(N$_c$) theory is periodic in the parameter $\theta=\mu_I/T$, 
owing to Z(N$_c$), and have observed that there are first-order vertical lines 
in the  $(T,\theta)$ plane, located at $\theta=2\pi(k+1/2)/N_c$, with integer $k$.
According to RW, these transitions are placed above a certain temperature 
$T_E$ and since the first of them appears for $\theta=\pi / N_c$, they conclude that
there is not transition at $\mu=0$. The last conclusion is contradicted 
in the literature~\cite{PDL} where there are different arguments according to 
which a chiral phase transition for $\mu=0$ does exist. Its nature depends on the 
number of quark, on fermion masses, and so on. 

\begin{figure}[tb]
\includegraphics[width=5.8cm,bb=74 448 530 743,clip]{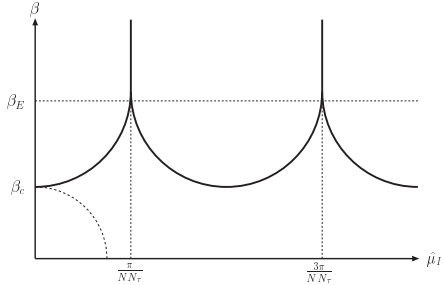}
\vspace{-0.5cm}
\caption[]{Schematization of the phase diagram in the $(\beta,\hat\mu_I)$ 
plane. The curved dotted line represents the critical line in the 
$(\beta,\hat\mu_R)$ plane.}
\label{fig0}
\vspace{-0.25cm}
\end{figure}

In view of this, the worst case which can occur, for the purposes of analytical 
continuation (AC), is that the RW critical line at $\theta=\pi / N_c$ could be 
prolongated downwards and bend up to reaching the $\theta=0$ axis in 
correspondence of a critical temperature $T_c$.
The RW periodicity and the $\mu\to-\mu$ symmetry, resulting from CP 
invariance, imply that knowing the expectation value of any observable 
for $T$ and $\theta$ inside the strip $\{0\leq \theta \leq \pi/N_c, 
\; 0\leq T < \infty \}$ in the $(T,\theta)$ plane is enough to fix it
on the remaining part of the plane. 
It is more suitable for lattice simulations to introduce the inverse 
coupling $\beta$ and the chemical potential in lattice units, 
$\hat \mu \equiv a \mu$. The RW transition lines in the $(\beta,\hat \mu_I)$ plane 
appear at $\hat\mu_I=2\pi(k+1/2)/(N_c N_{\tau})$ and 
we can consider only the strip $\{0\leq \hat\mu_I \leq \pi/(N_c N_{\tau});  
0\leq \beta < \infty \}$ (see Fig.~\ref{fig0}).

\begin{figure}[tb]
\includegraphics[width=5.7cm,angle=-90]{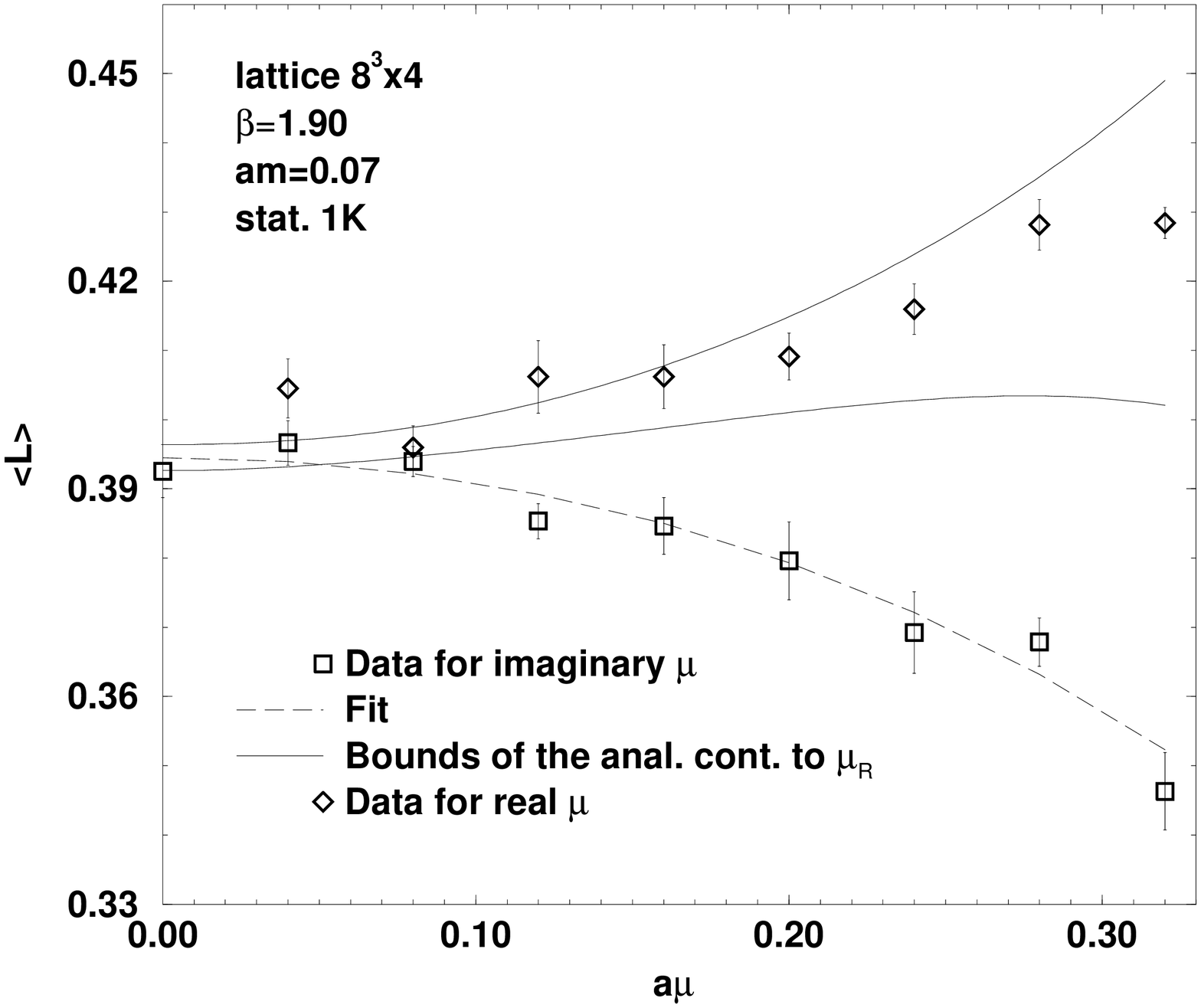}
\includegraphics[width=5.7cm,angle=-90]{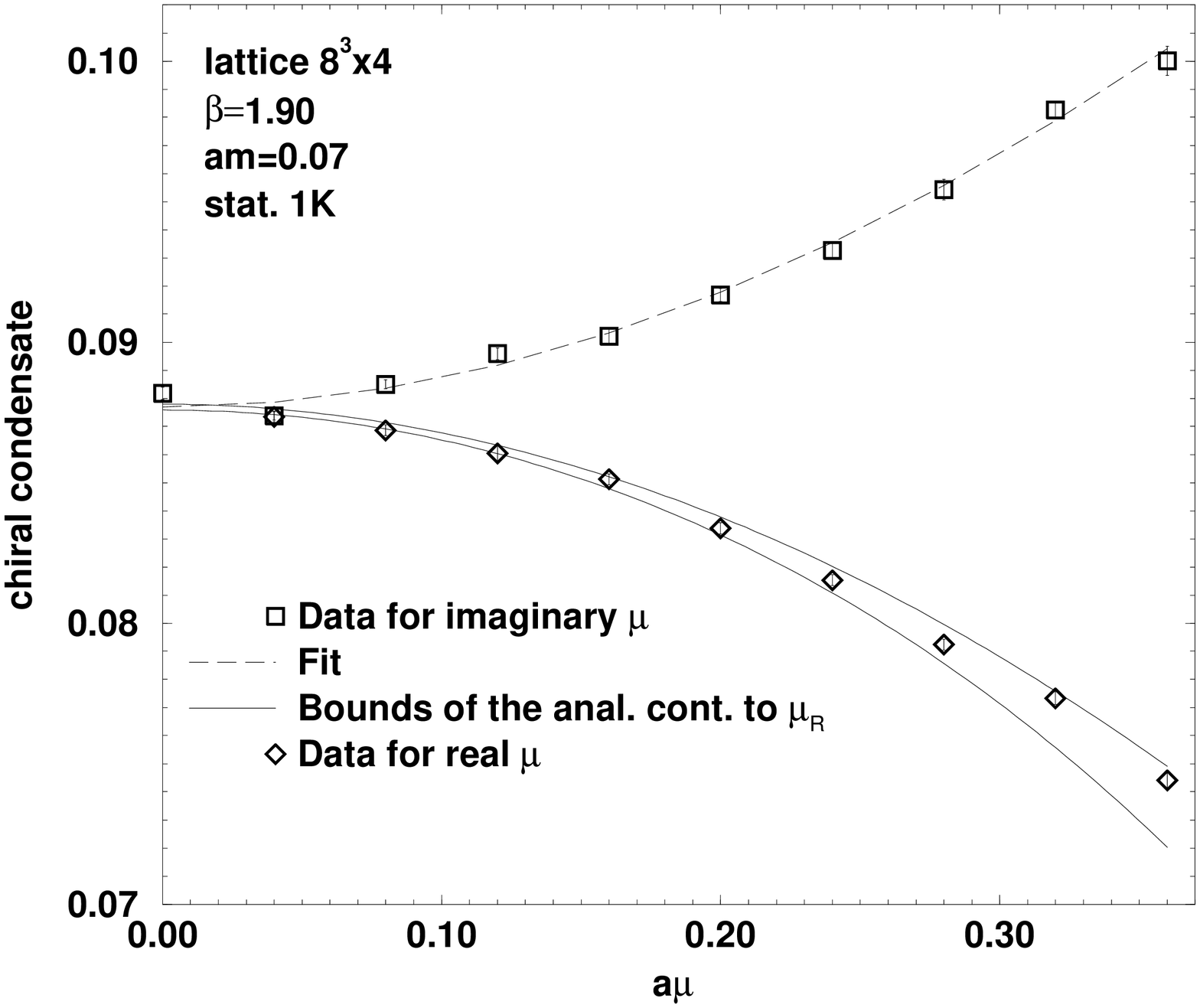}
\vspace{-0.5cm}
\caption[]{Polyakov loop (above) and chiral condensate (below) {\it vs}
$\hat \mu$ for $\beta=1.90$.} 
\label{fig1}
\vspace{-0.25cm}
\end{figure}

\begin{figure}[tb]
\includegraphics[width=5.7cm,angle=-90]{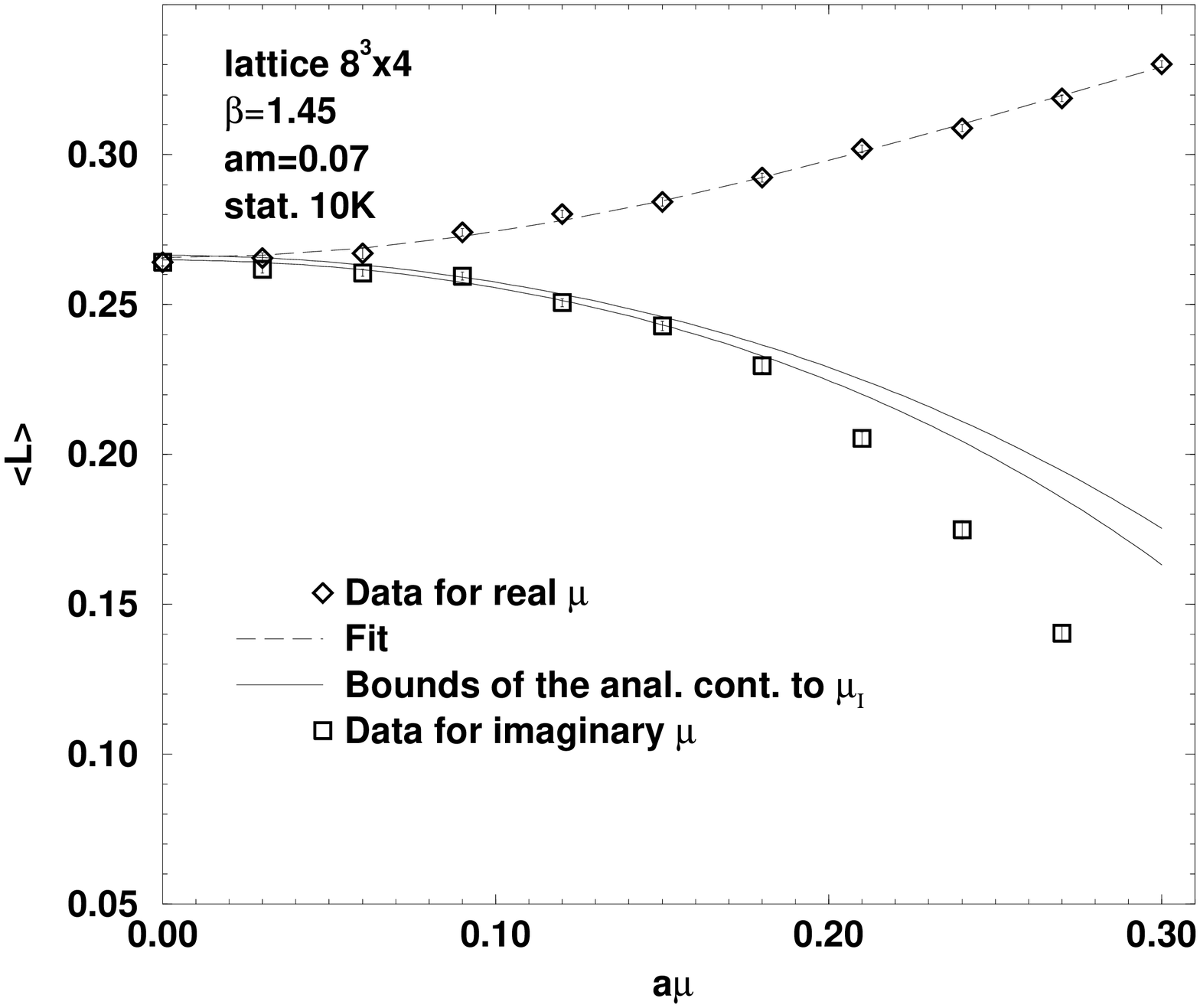}
\includegraphics[width=5.7cm,angle=-90]{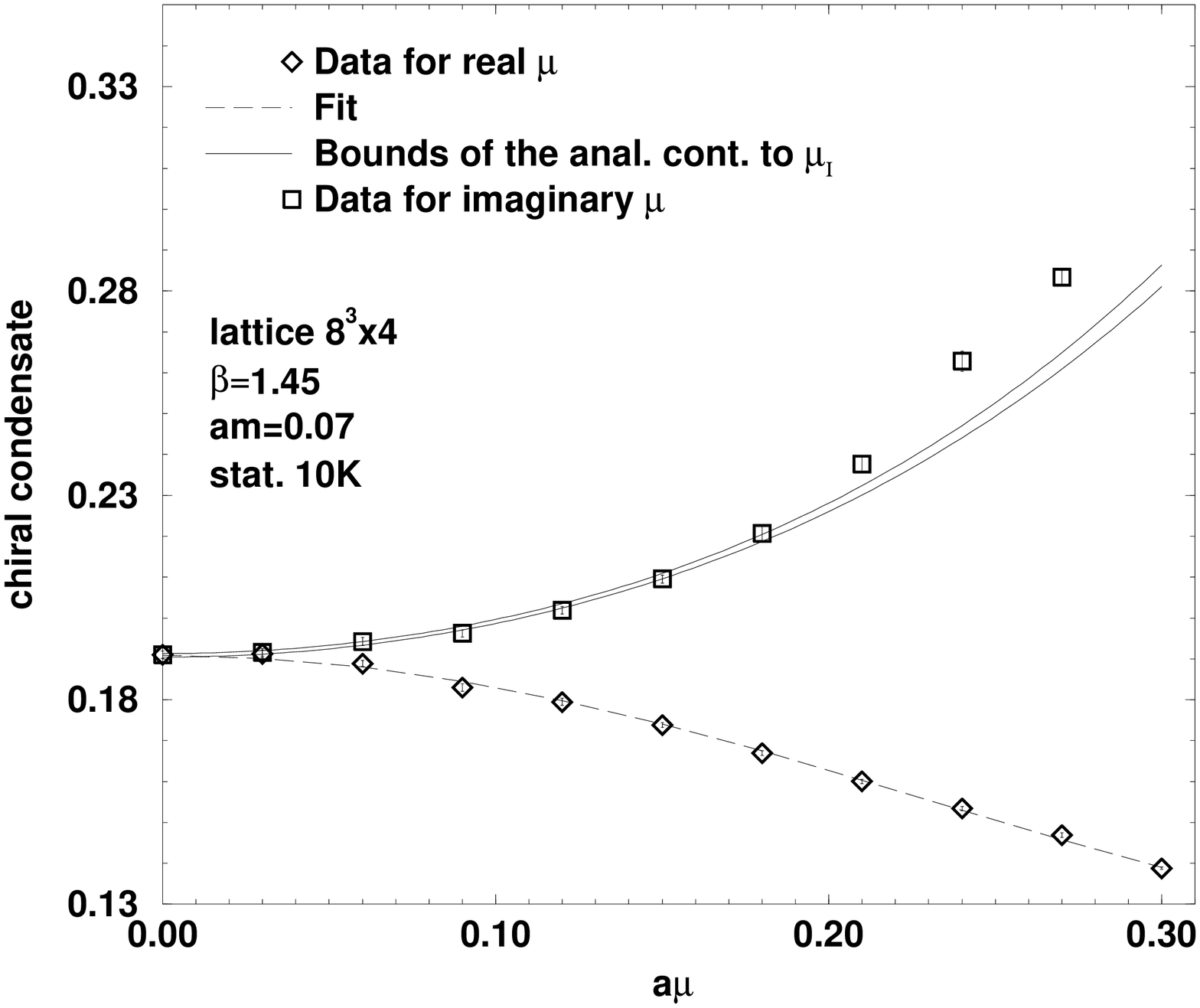}
\vspace{-0.5cm}
\caption[]{Polyakov loop (above) and chiral condensate (below) {\it vs}
$\hat \mu$ for $\beta=1.45$.}
\label{fig2}
\end{figure}

\section{NUMERICAL RESULT}
In this work we consider SU(2) gauge theory with staggered fermions ($n_f=8$)
because in this case numerical simulations are feasible 
for both real and imaginary chemical potential, thus allowing to test the 
method of AC by direct comparison.
We performed all simulations on a $8^3 \times 4$ lattice, setting the quark 
bare mass to $am=0.07$.
For different values of $\beta$ we performed simulations for both real 
and imaginary chemical potential, varying $\hat \mu_I$ and $\hat \mu_R$ 
between 0 and $\pi/(2 N_\tau)$ and determined expectation values of    
the Polyakov loop $L$ and the chiral condensate $\langle \bar \psi \psi 
\rangle$.
Then, we interpolated data obtained for imaginary $\mu$
with a {\em truncated} Taylor series of the form {$a+b\,\hat \mu_I^2+c\,\hat
\mu_I^4$}. 
After ``rotating'' this polynomial to real chemical potential, thus leading to
$a-b\,\hat\mu_R^2+c\,\hat\mu_R^4$, we compared it with the 
determinations obtained directly for real $\mu=\mu_R$.

We selected three values of $\beta$ (1.90, 1.45, 0.90), each representative of 
one of the regions $\beta>\beta_E$, $\beta_c<\beta<\beta_E$ and 
$\beta<\beta_c$. 

For $\beta=1.90$ we interpolated the data obtained for 
$0\leq\hat \mu_I  \lesssim \pi/(2N_\tau)$
and found (see Fig.~\ref{fig1}) that the rotated polynomial perfectly 
interpolates data obtained directly for real $\mu=\mu_R$ in the same range. 
From this outcome we can conclude that the AC {works} 
in the region $\beta>\beta_E$ for all the $\hat\mu_I$'s before the first RW 
critical line.

For $\beta=1.45$ we are in the region $\beta_c < \beta <\beta_E$ for which
we expect that by varying $\hat \mu_I$ the chiral critical line is crossed, while
no transitions should be met at the same $\beta$ by varying $\hat \mu_R$.
Therefore this time we interpolated data obtained for the {\em real} chemical 
potential $0\leq\hat\mu_R \lesssim\pi/(2N_\tau)$ and rotated this polynomial to its
counterpart in $\hat\mu_I$. The comparison (see Fig.~\ref{fig2}) with data 
obtained directly for imaginary chemical potential shows agreement for 
$\hat\mu_I$ below {$\simeq 0.18$}, which can be taken as an estimate of the 
chiral critical value in the $(\beta,\hat\mu_I)$ plane at the given $\beta$. 

\begin{figure}[tb]
\includegraphics[width=5.7cm,angle=-90]{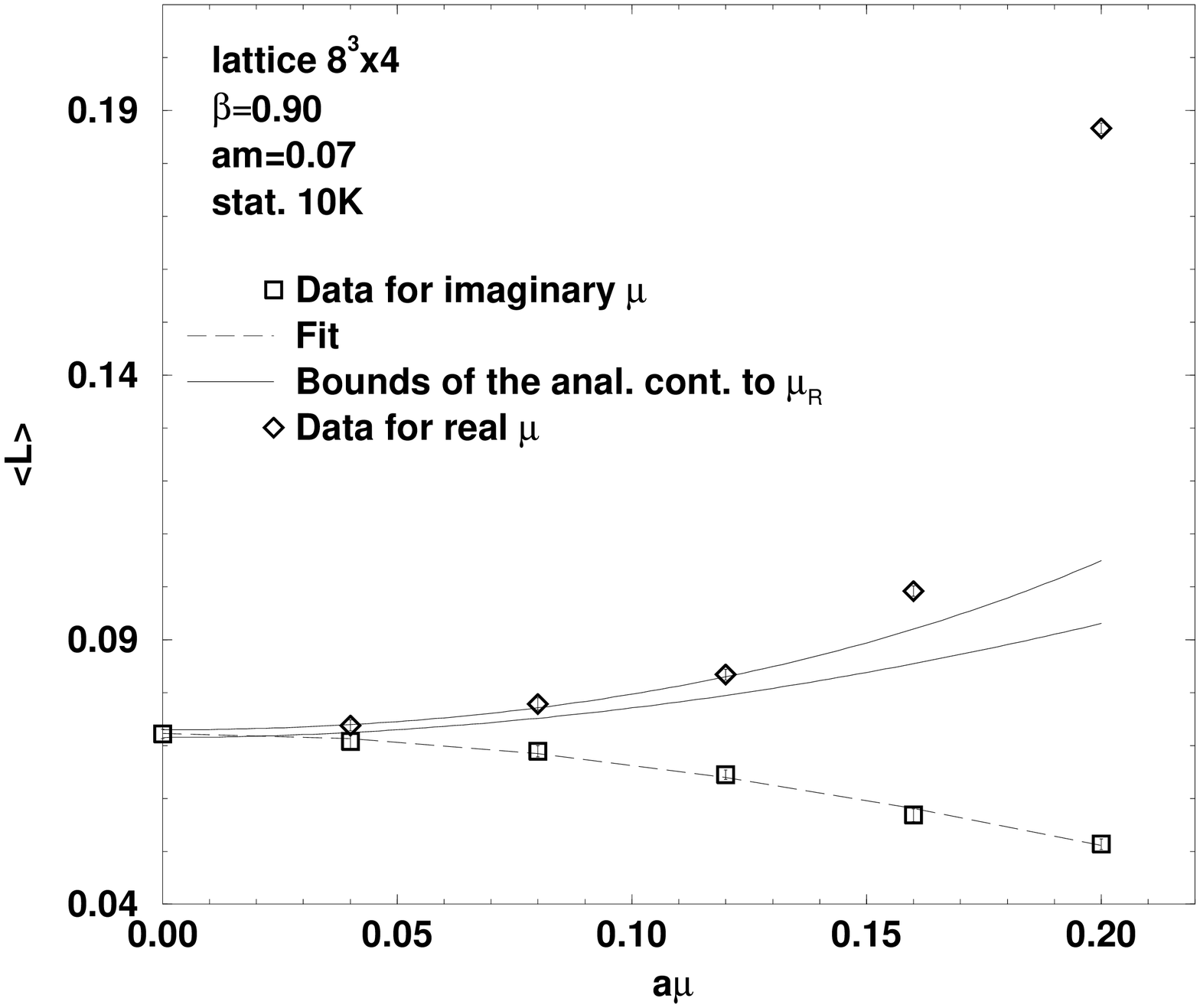}
\includegraphics[width=5.7cm,angle=-90]{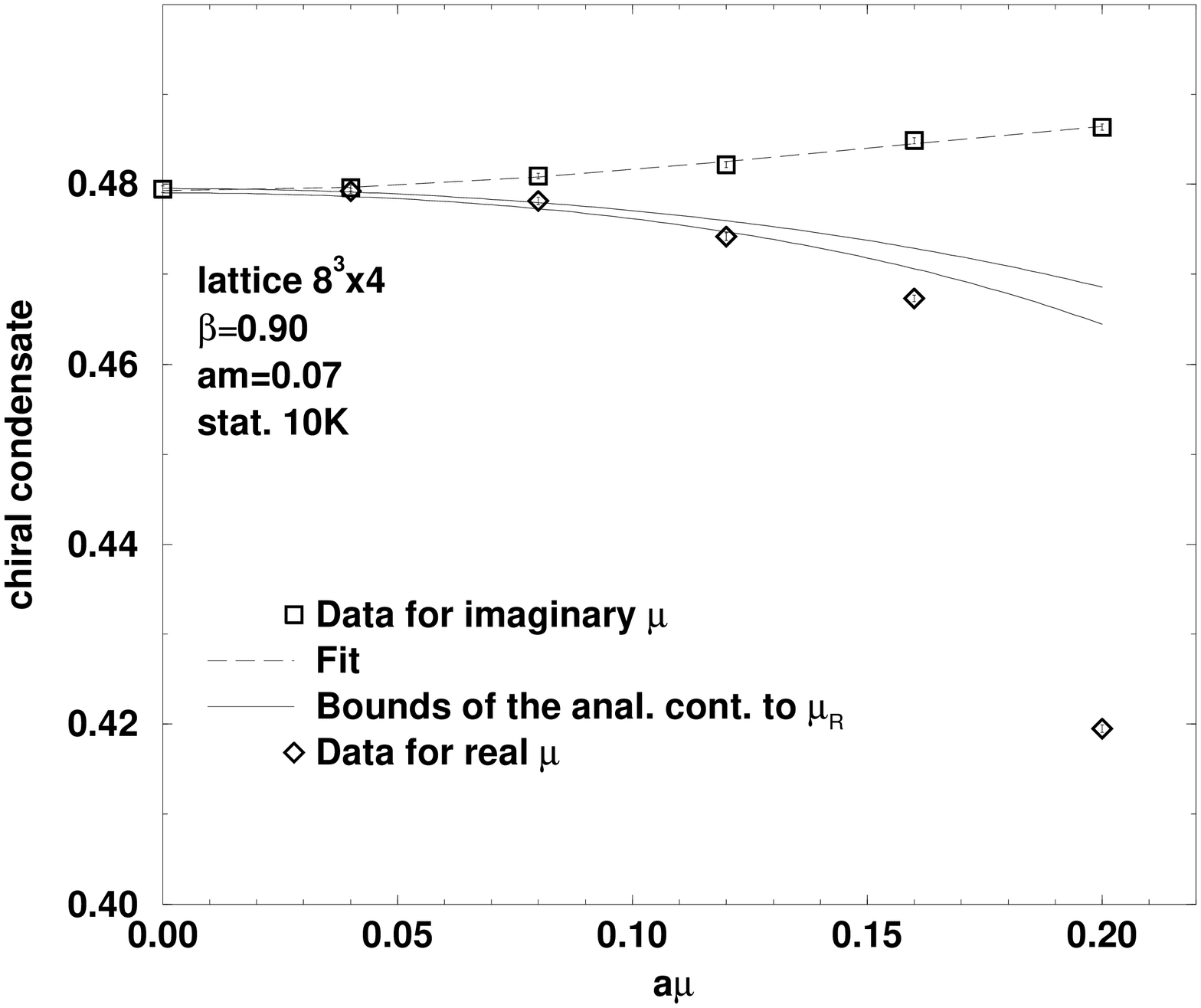}
\vspace{-0.5cm}
\caption[]{Polyakov loop (above) and chiral condensate (below) {\it vs}
$\hat \mu$ for $\beta=0.90$.} 
\label{fig3}
\end{figure}

For $\beta=0.90<\beta_c$, we expect analyticity for all possible
values of $\hat \mu_I$, while the chiral critical line in the 
$(\beta,\hat\mu_R)$ plane is crossed by varying $\hat\mu_R$ at the given  
value of $\beta$. Therefore we interpolated data obtained for 
$0\leq\hat\mu_I\lesssim\pi/(2N_\tau)$ and rotated this polynomial to its  
counterpart in $\hat\mu_R$.
The comparison (see Fig.~\ref{fig3}) with data obtained directly for real 
chemical potential shows agreement for $\hat\mu_R$ {below} {$\simeq 0.12$}, 
which can be taken as an estimate of the critical value in the 
$(\beta,\hat\mu_R)$ plane at the given $\beta$. 

\section{CONCLUSION}
We have presented a possible structure of the phase diagram 
in the $(T,\mu_I)$ plane for 2--color QCD with $n_f=8$ and have found that 
the method of AC works fine within the restrictions imposed by the presence 
of nonanalyticities.

\end{document}